\documentclass[preprint,12pt]{elsarticle}
\usepackage{}

\usepackage{amsmath}
\usepackage{cases}
\usepackage{hyperref}
\usepackage{amssymb}
\usepackage{graphicx}
\usepackage{subfigure}
\usepackage{slashbox}
\usepackage{pict2e}

\biboptions{numbers,sort&compress}

\begin{document}
\begin{frontmatter}

%% Title, authors and addresses

%% use the tnoteref command within \title for footnotes;
%% use the tnotetext command for the associated footnote;
%% use the fnref command within \author or \address for footnotes;
%% use the fntext command for the associated footnote;
%% use the corref command within \author for corresponding author footnotes;
%% use the cortext command for the associated footnote;
%% use the ead command for the email address,
%% and the form \ead[url] for the home page:
%%
%% \title{Title\tnoteref{label1}}
%% \tnotetext[label1]{}
%% \author{Name\corref{cor1}\fnref{label2}}
%% \ead{email address}
%% \ead[url]{home page}
%% \fntext[label2]{}
%% \cortext[cor1]{}
%% \address{Address\fnref{label3}}
%% \fntext[label3]{}

\title{The influence of numerical noises \\ on  statistics computation of chaotic dynamic systems}

%% use optional labels to link authors explicitly to addresses:
%% \author[label1,label2]{<author name>}
%% \address[label1]{<address>}
%% \address[label2]{<address>}

\author{Xiaoming Li}
\author{Shijun Liao\corref{cor1}}
\ead{sjliao@sjtu.edu.cn} \cortext[cor1]{Corresponding author.}

\address{State Key Laboratory of Ocean Engineering, Shanghai 200240, China}

\address{MoE Key Laboratory in Scientific and Engineering Computing, Shanghai 200240, China}

\address{School of Naval Architecture, Ocean and Civil Engineering,   
Shanghai Jiao Tong University, Shanghai 200240, China}

\begin{abstract}
It is well known that chaotic dynamic systems (such as three-body system, turbulent flow and so on) have the sensitive dependance on initial conditions (SDIC).   Unfortunately, numerical noises (such as truncation error and round-off error) always exist in practice.  Thus, due to the SDIC, long-term accurate prediction  of  chaotic dynamic systems is practically impossible, and therefore numerical simulations of chaos are only mixtures of ``true'' solution with physical meanings and numerical noises without physical meanings.  However,  it is traditionally believed that statistic computations based on such kind of ``mixtures'' of numerical simulations of chaotic dynamic systems are acceptable.   In this paper, using the so-called ``clean numerical simulation'' (CNS) whose numerical noises might be much smaller even than micro-level physical uncertainty and thus are negligible, we gain accurate prediction of a chaotic dynamic system in a long enough interval of time.   Then, based on these reliable simulations,  the influence of numerical noises on statistic computations is investigated.  It is found that the influence of numerical noises is negligible when statistic results are time-independent.  Unfortunately, when a chaotic dynamic system is far from equilibrium state so that its statistics are time-dependent, numerical noises have a great influence even on statistic computations.  It suggests that even the direct numerical simulations (DNS) might not give reliable statistic computations for non-equilibrium dynamic systems with the SDIC.
\end{abstract}

\begin{keyword}
%% keywords here, in the form: keyword \sep keyword

%% MSC codes here, in the form: \MSC code \sep code
%% or \MSC[2008] code \sep code (2000 is the default)
chaos \sep numerical noises \sep statistics  \sep Clean Numerical Simulation (CNS) 
\end{keyword}

\end{frontmatter}

%%
%% Start line numbering here if you want
%%
% \linenumbers

%% main text
\section{Introduction}

It is well known that chaotic dynamic systems (such as three-body system) have the sensitive dependance on initial conditions (SDIC) \cite{Lorenz1963}, which is called the ``butterfly-effect''.    Unfortunately, numerical noises, i.e.  truncation error and round-off error,  always exist in practice.  Thus, due to the SDIC, long-term accurate prediction  of  chaotic dynamic systems is practically impossible \cite{Lorenz1963} and therefore numerical simulations of chaos are only mixtures of ``true'' solution with physical meanings and numerical noises {\em without} physical meanings:  the latter is often at the same level as the former, due to the exponential growth of the initial tiny uncertainty of chaotic dynamic system.    Note that it is widely believed that turbulence \cite{Lee2015, wang2015study,Avila2011, Deike2014} is chaotic.   So,  statistics are commonly used in the study of turbulence, and the direct numerical simulation (DNS) \cite{KMM1987},  which solves the Navier-Stokes equations without averaging or approximation but with all essential scales of motion, has played an important role in statistical computation of turbulence.   However, due to the SDIC, tiny numerical noises, which grow  exponentially in time,  may lead to spurious results. Thus,  DNS results are  in  essence  the mixtures of ``true'' solution with physical meaning and spurious results {\em without} physical meanings:  it is even worse that the latter is often at the same level as the former and besides it is very hard to  seperate the former from the whole.   Even so,  it is widely believed that numerical noises do not affect the statistics of turbulence by means of such kind of unreliable  simulations.  Unfortunately,  Yee et al.  \cite{Yee1999}  reported   in 1999  that the DNS can produce a spurious solution that is completely different from the  physical solution of their considered equations.   In addition,  Wang et al. \cite{Wang2009} demonstrated a spurious evolution of turbulence originated from round-off error in DNS.  For more examples of spurious numerical simulations, please refer to  Yee et al. \cite{Yee1991, Yee1995}.   Therefore, it is necessary to verify  the reliability of statistic results for chaotic dynamic systems.

Recently,  a  numerical  method with extremely small numerical noises,  namely ``clean numerical simulation" (CNS) \cite{Liao2009}, was  proposed  to  gain  reliable  simulations  of  chaotic dynamic systems in  a  finite but large  enough  interval  of  time.   The  CNS  is  based  on an arbitrary Taylor series method (TSM) \cite{Corliss1982, Barrio2005} and an arbitrary multiple precision  \cite{Oyanarte1990} for every data,  together with a kind of solution verification.   By means of the CNS, the numerical noises can be  so greatly reduced    to  be much smaller than the  ``true''  solution that the numerical noises are  negligible  in a given interval of time even  for  chaotic  dynamic  systems.   The  CNS  has been successfully applied in some chaotic dynamic systems, such as the famous  Lorenz equation \cite{Liao2009,Wang2012, Liao-Wang2014, Liao2014} and the three-body problems \cite{Liao2013-3b, Li-Liao2014, Liao2015}.     Note that, using the traditional Runge-Kutta method and  data in double precision, one can gain convergent chaotic results of Lorenz equation only in a few dozens of time interval.   However,  using the CNS,  Liao  and  Wang \cite{Liao-Wang2014} successfully obtained a convergent, reliable chaotic solution of the Lorenz equation in an interval [0,10000], which is hundreds times  larger than those given by the traditional numerical methods.  This illustrates the validity of the CNS  for  reliable  simulations  of  chaotic dynamic systems with the SDIC.

\section{Mathematical equations}  %%

It is well known that the famous Lorenz equation \cite{Lorenz1963} is a very simplified model of the Rayleigh-B\'{e}nard (RB) flow of viscous fluid.  From the  exact  Navier-Stokes equations for the Rayleigh-B\'{e}nard flow 
\begin{eqnarray}
    \nabla^2 \frac{\partial{\psi}}{\partial t} + \frac{\partial{(\psi, \nabla^2\psi)}}{\partial(x,z)} - \sigma \frac{\partial\theta}{\partial x} -
    \sigma \nabla^4\psi = 0 \\
    \frac{\partial\theta}{\partial t} + \frac{\partial(\psi,\theta)}{\partial(x,z)} - R\frac{\partial\psi}{\partial x} - \nabla^2\theta = 0
\end{eqnarray}
where $\psi$ denotes the stream function, $\theta$ the temperature  departure  from a linear variation background,  $t$ the time, $(x,z)$ the horizontal and vertical coordinates, $\sigma$ the Prandtl number, $R$ the Rayleigh number, respectively,  Saltzman \cite{saltzman1962finite}  deduced  a  family of highly  truncated  dynamic systems with different degrees of freedom,  and the famous Lorenz equation \cite{Lorenz1963} is only the simplest  one  among  them.    For example,   in the case of the Prandtl number $\sigma=10$, the highly truncated dynamic system with three degrees of freedom  (3-DOF) reads
\begin{equation}\label{3eqns}
\left\{
\begin{split}
  \dot A &= -148.046A - 1.500D, \\
  \dot D &= -13.958AG - 1460.631\lambda A - 14.805D, \\
  \dot G &= 27.916AD - 39.479G,
\end{split}
\right.
\end{equation}
where $\lambda=R/R_c$ is dimensionless Rayleigh number, $R_c$ is the critical Rayleigh number,  $A$ and $D$ represent the cellular streamline and thermal fields for the Rayleigh critical mode, and $G$ denotes the departure of the vertical variation, respectively.  For details, please refer to  Saltzman \cite{saltzman1962finite}.   

Similarly, Saltzman \cite{saltzman1962finite}  gave the highly truncated dynamic system with five degrees of freedom  (5-DOF):
\begin{equation}\label{5eqns}
\left\{
\begin{split}
  \dot A &= 23.521BC - 1.500D - 148.046A, \\
  \dot B &= -22.030AC - 186.429B, \\
  \dot C &= 1.561AB - 400.276C, \\
  \dot D &= -13.958AG - 1460.631\lambda A - 14.805D, \\
  \dot G &= 27.916AD - 39.479G,
\end{split}
\right.
\end{equation}
and that with the seven degrees of freedom (7-DOF):
\begin{equation}\label{7eqns}
\left\{
\begin{split}
    \dot A &= 23.521BC - 1.500D - 148.046A, \\
    \dot B &= -22.030AC - 1.589E - 186.429B, \\
    \dot C &= 1.561AB - 0.185F - 400.276C, \\
    \dot D &= -16.284CE - 16.284BF - 13.958AG \\
           &\qquad\qquad\qquad   - 1460.631\lambda A -14.805D, \\
    \dot E &= 16.284CD - 16.284AF - 18.610BG \\
           &\qquad\qquad\qquad   -1947.508\lambda B -18.643E, \\
    \dot F &= 16.284AE + 16.284BD - 486.877\lambda C\\
             & \qquad\qquad\qquad  - 40.028F, \\
    \dot G &= 27.916AD + 37.220BE - 39.479G,
\end{split}
\right.
\end{equation}
respectively.   All of them are deterministic equations with chaotic solutions, and are greatly simplified models for the two dimensional Rayleigh-B\'{e}nard flow.   It should be emphasized  here that, for any a given initial condition, we can gain reliable, convergent numerical results of chaotic solutions of these models in a finite but  long enough interval of time by means of the CNS \cite{Liao2009,Liao-Wang2014, Liao2014}.

In physics, the Rayleigh-B\'{e}nard flow with  a large  enough Rayleigh number $R$ is an evolutionary process from an initial equilibrium state  to  turbulence after a long enough time, mainly because the flow is unstable and besides the micro-level physical uncertainty (such as thermal fluctuation) always exists.   Such kind of initial micro-level physical uncertainty due to thermal fluctuation can be expressed by Gaussian random data, as illustrated by Wang et al.  \cite{wang2015study}.   Mathematically, due to the SDIC,  the reliable, convergent chaotic solutions of these simplified models should be dependent upon the random initial conditions.   This is true in physics,  since each experimental observation of the Rayleigh-B\'{e}nard flow is different.    Therefore,  these equations provide us some  simplified  models to investigate the influence of numerical errors on the statistics computation of such kind of random process.

Without loss of generality, let us first  consider the deterministic 3-DOF equations (\ref{3eqns}) with the random initial conditions in  normal  distribution  in case of $\lambda = 28$,  corresponding to a turbulent flow.   Considering the thermal fluctuation,  we study here such kind of random  initial  condition in normal distribution  with the mean
\begin{displaymath}
   \langle A(0) \rangle =1, \quad  \langle D(0) \rangle = 10^{-3},\quad   \langle G(0) \rangle = 10^{-3}
\end{displaymath}
and the standard deviation
\begin{equation}
\sigma_{0}  = \sqrt{\langle  A^2(0)\rangle}=\sqrt{\langle D^2(0)\rangle}=\sqrt{\langle G^2(0)\rangle} =10^{-30}.   \nonumber
\end{equation}
By means of the CNS, we can gain reliable propagations of the micro-level physical uncertainty of a large numbers of random initial conditions.   Let
\begin{eqnarray}
\langle A(t)\rangle &=&\frac{1}{N}\sum_{i=1}^{N}A_i(t),\\
\sigma_A(t) &=&\sqrt{\frac{1}{N-1}\sum_{i=1}^N[A_i(t)-\langle A(t)\rangle ]^2}
\end{eqnarray}
denote the sample mean and unbiased estimate of standard deviation of $A(t)$ of these reliable simultions,  respectively, where $N$ is the number of samples.

\section{Statistical Results}

\begin{figure}[h]
  \centering
  \includegraphics[scale=0.45]{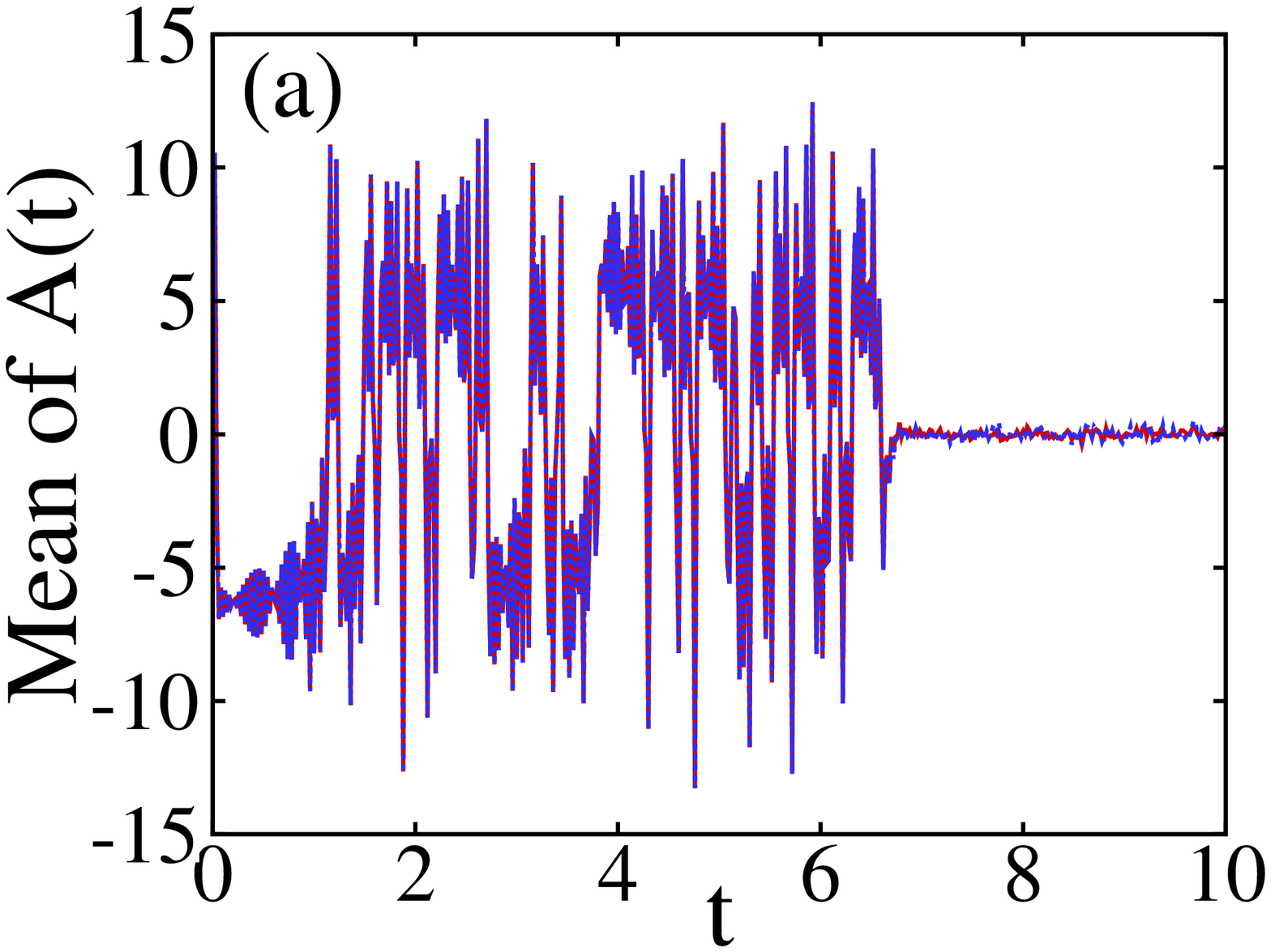}
  
  \includegraphics[scale=0.45]{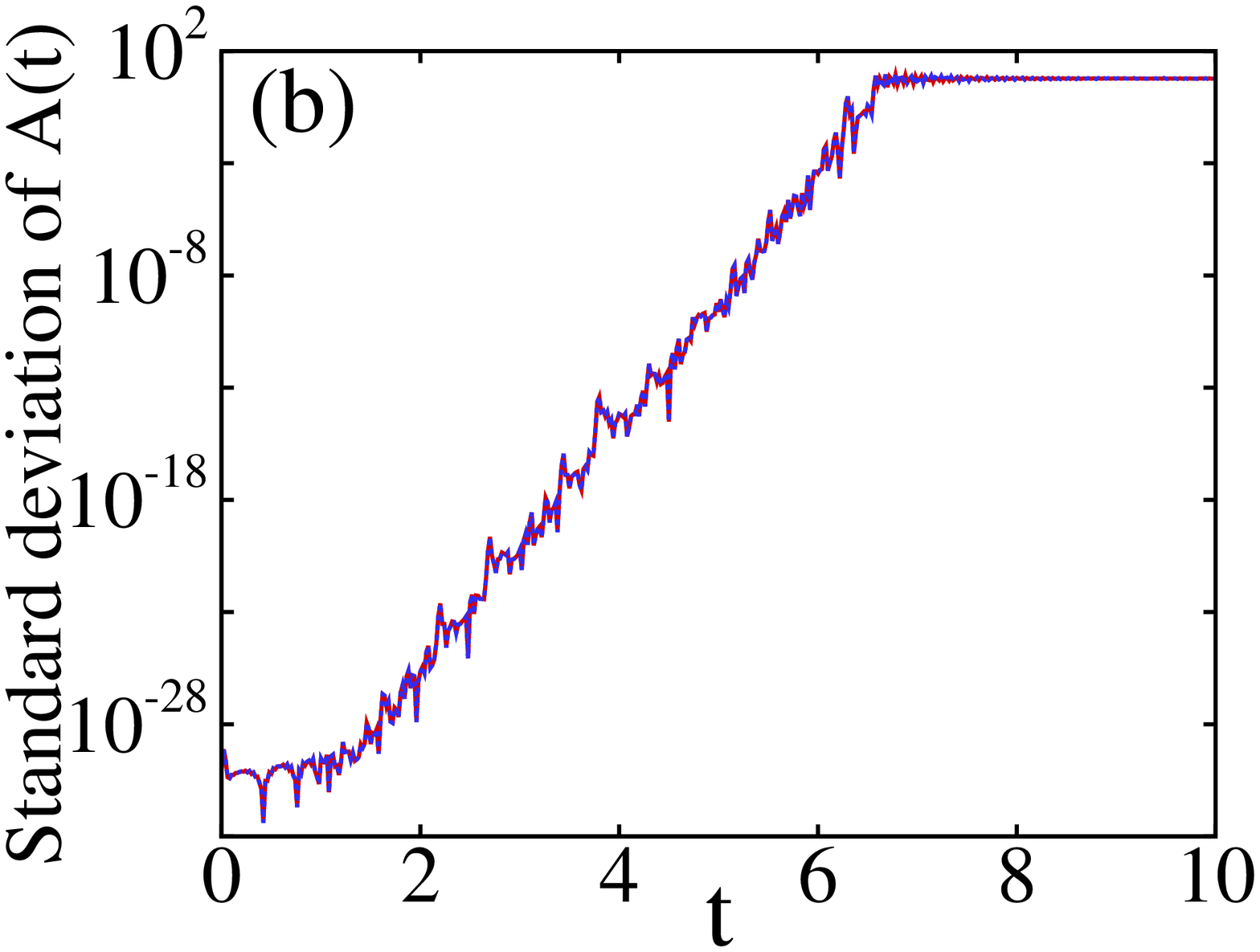}
   \caption{(a) The mean and (b) standard deviation of $A(t)$ of the 3-DOF chaotic system   (\ref{3eqns})  using different numbers of samples gained by means of the CNS.  Solid line: 2000 samples; Dashed line: 1000 samples.}
  \label{fig:3eqns-M80K300-2k-1k}
\end{figure}

Two thousand samples of reliable numerical simulations of the 3-DOF system (\ref{3eqns}) are obtained in the time interval $[0,10]$ by means of  the CNS using the 80th-order Taylor series ($M=80$),  the 90 decimal-digit precision ($K=90$)  for every data, and the time-step $\delta t =10^{-3}$.   It is found that the numerical errors can be  decreased to be much smaller than the micro-level physical uncertainty in the time interval [0,10] under consideration.    These numerical simulations are so accurate that we can consider them as the ``true'' solutions of the chaotic dynamic system (\ref{3eqns}), which can be used to investigate the influence of numerical noises on statistic computations of chaotic systems.    In this way,  we successfully distinguish the ``true'' chaotic solutions with physical meanings from the numerical noises {\em without} physical meanings!   Note that, for the chaotic dynamic system (\ref{3eqns}),  the mean and the standard deviation of $A(t)$ using 1000 samples  are almost the same as those using 2000 samples, as shown in Figure \ref{fig:3eqns-M80K300-2k-1k}.  Thus, it is enough for us to use 2000 samples in this paper.

\begin{figure}[h]
  \centering
  \includegraphics[scale=0.45]{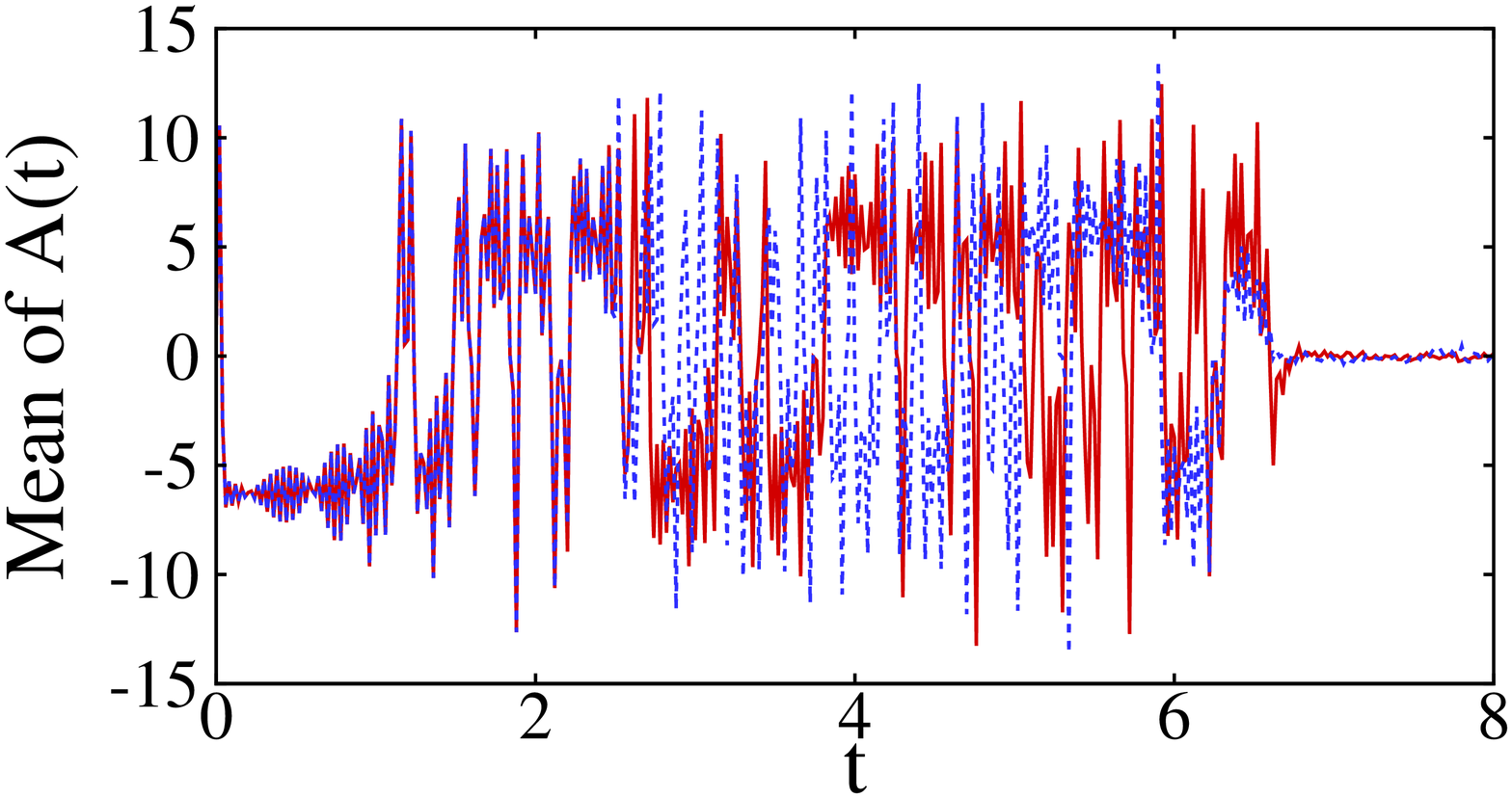}
  \caption{The influence of the truncation error on the mean of $A(t)$ of the 3-DOF chaotic dynamic system (\ref{3eqns}) using the time-step $\delta t =10^{-3}$  and  the 90 decimal-digit precision for every data (i.e. with the negligible round-off error).   Solid line in red:  the reliable  results given by the CNS using $M=80$;  dashed line in blue:  the result given by $M=10$.}
  \label{fig:3eqns-M80K300-M10K300}
\end{figure}

Obviously, the larger the order $M$ of the Taylor series in the frame of the CNS, the smaller the truncation error.   For example,  the traditional Runge-Kutta method corresponds to $M=4$, the 4th-order Taylor series expansion for $t$.  Thus, to investigate the influence of the truncation error,  we use here the 10th-order Taylor series, i.e. $M=10$,  but retain the 90 decimal-digit multiple precision for every data.     Note that our reliable CNS results are gained by means of $M=80$, i.e. the 80th-order Taylor series, whose truncation errors are negligible in the considered interval of time $t\in[0,10]$.  However,  when $M=10$,  the  truncation  error is {\em not}  negligible in [0,10].   In this way,  the round-off error is negligible in the considered interval of time $t\in[0,10]$ so that the influence of the truncation error can be investigated independently.

\begin{figure}[th]
  \centering
  \includegraphics[scale=1.0]{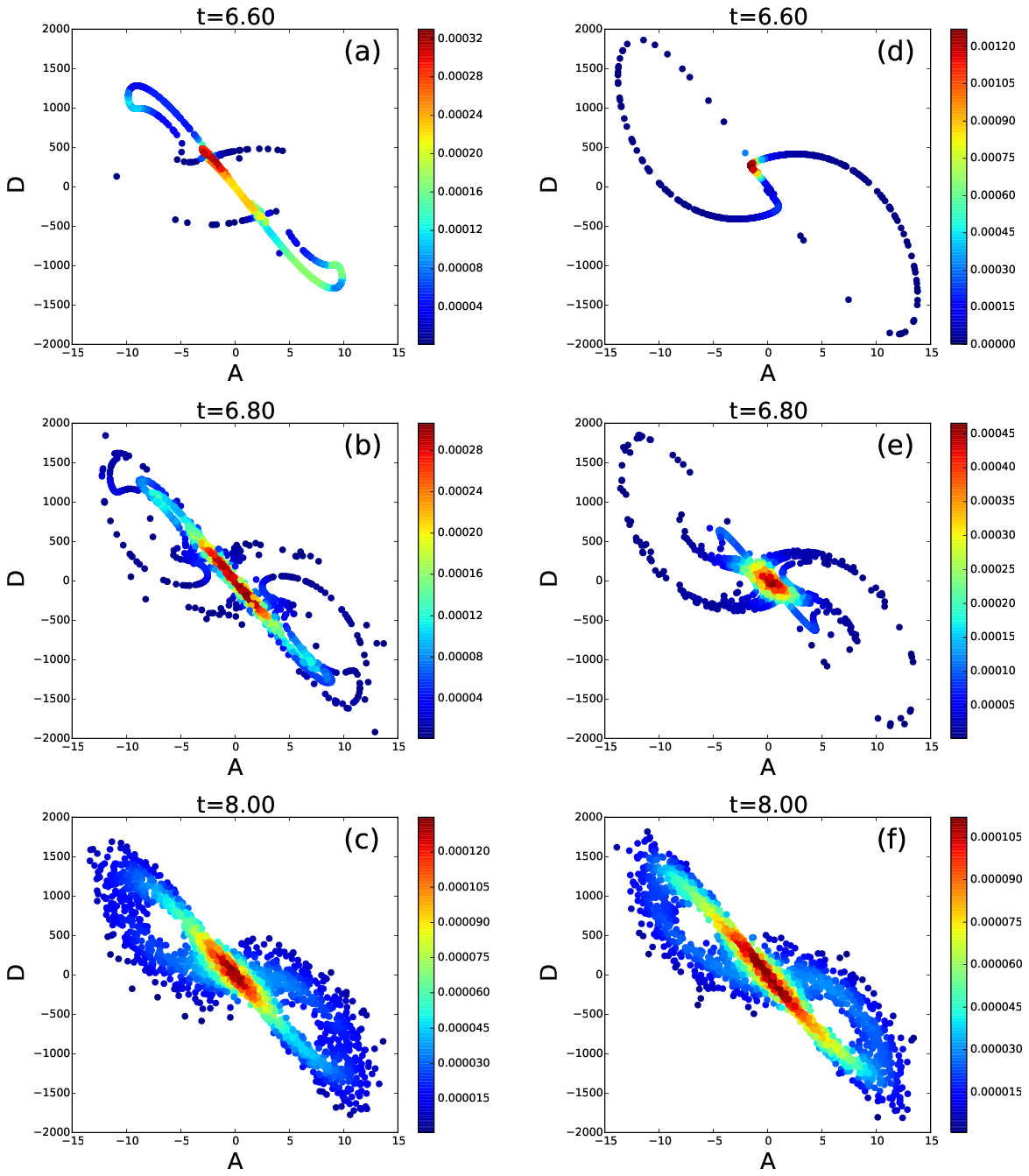}
  \caption{The scatter diagrams of the probability distribution function (PDF) in the $A - D$ plane of the 3-DOF system (\ref{3eqns}) at different times by means of the time-step $\delta t =10^{-3}$,  the 90 decimal-digit precision for every data and the different orders $M$ of Taylor series.  (a)-(c): unreliable results given by $M=10$; (d)-(f): the reliable  results given by the CNS using $M=80$.}
  \label{fig:3eqns-GaussianKDE}
\end{figure}

The reliable mean  (the red line, given by $M=60$)  of $A(t)$ of the 3-DOF chaotic dynamic system (\ref{3eqns}) is as shown in Figure~\ref{fig:3eqns-M80K300-M10K300}, compared with the unreliable result (the blue line, given by $M=10$).    Note that the reliable mean of $A(t)$ given by the CNS becomes stable when $t>7$, but is time-dependent when $t\leq 7$.  In physics, it means that the chaotic 3-DOF dynamic system is unstable when $t \leq 7$, even from the view-point of statistics, since the RB flow is in a transition stage from an equilibrium state to a new equilibrium one.   However,  as shown in Figure~\ref{fig:3eqns-M80K300-M10K300}, the unreliable mean of $A(t)$  given by $M=10$ has a noticeable difference from the reliable CNS result in  the time interval $2.5<t<7$.
It suggests that the truncation error has a great influence on the computation of unsteady statistical quantities when the chaotic dynamic system is in a transition stage from an equilibrium state to a new equilibrium one.   
Note that the mean of $A(t)$ given by $M=10$ agrees well with the reliable CNS result when $t>7$.  Thus, the truncation error has no influence on the time-independent statistics of chaotic dynamic systems in an equilibrium state.  This might be the reason why many DNS results for fully developed turbulence agree well with experimental data.   

\begin{figure}[h]
  \centering
  \includegraphics[scale=0.45]{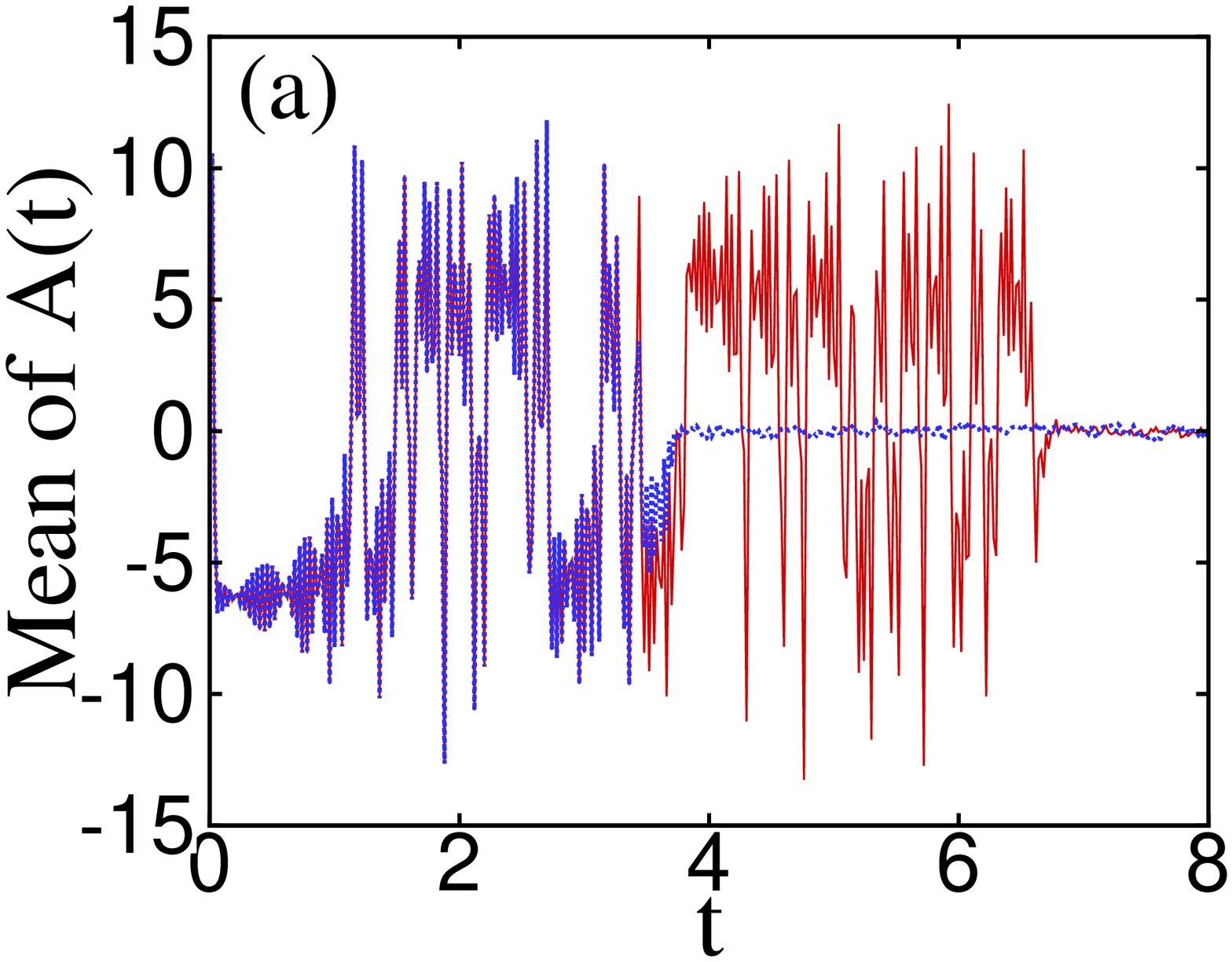}
  
  \includegraphics[scale=0.45]{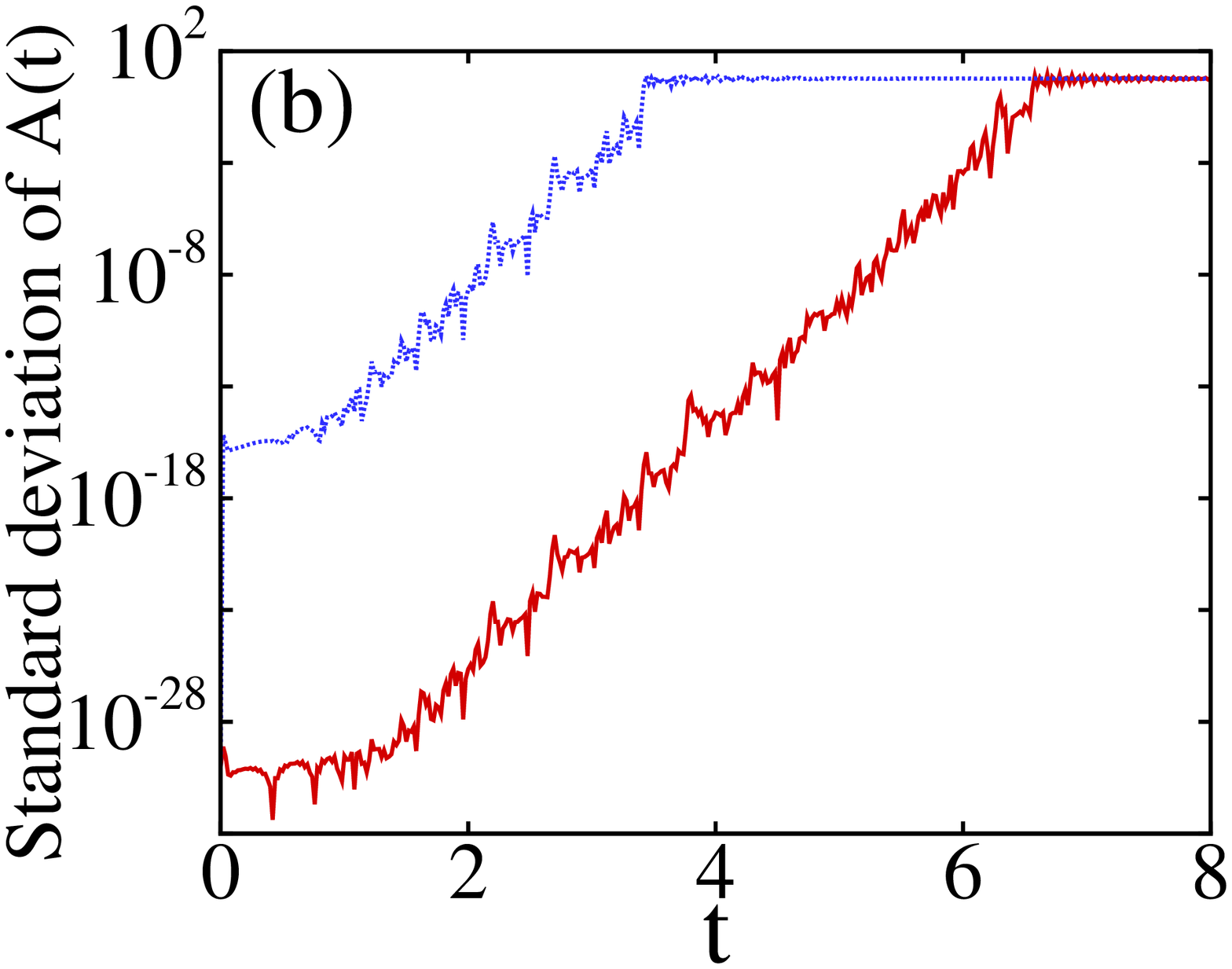}
  \caption{The influence of the round-off error on (a) the mean and (b) the standard deviation of $A(t)$ of the 3-DOF chaotic system (\ref{3eqns})  using the time-step $\delta t= 10^{-3}$ and the 80th-order Taylor expansion ($M=80$) with the negligible truncation error.   Solid line in red: the reliable CNS results given by the 90 decimal-digit multiple precision for every data; dashed line in blue:  the unreliable results given by means of double precision.}
\label{fig:3eqs-M80K300-roundoff}
\end{figure}

\begin{figure}[h]
  \centering
  \includegraphics[scale=1.0]{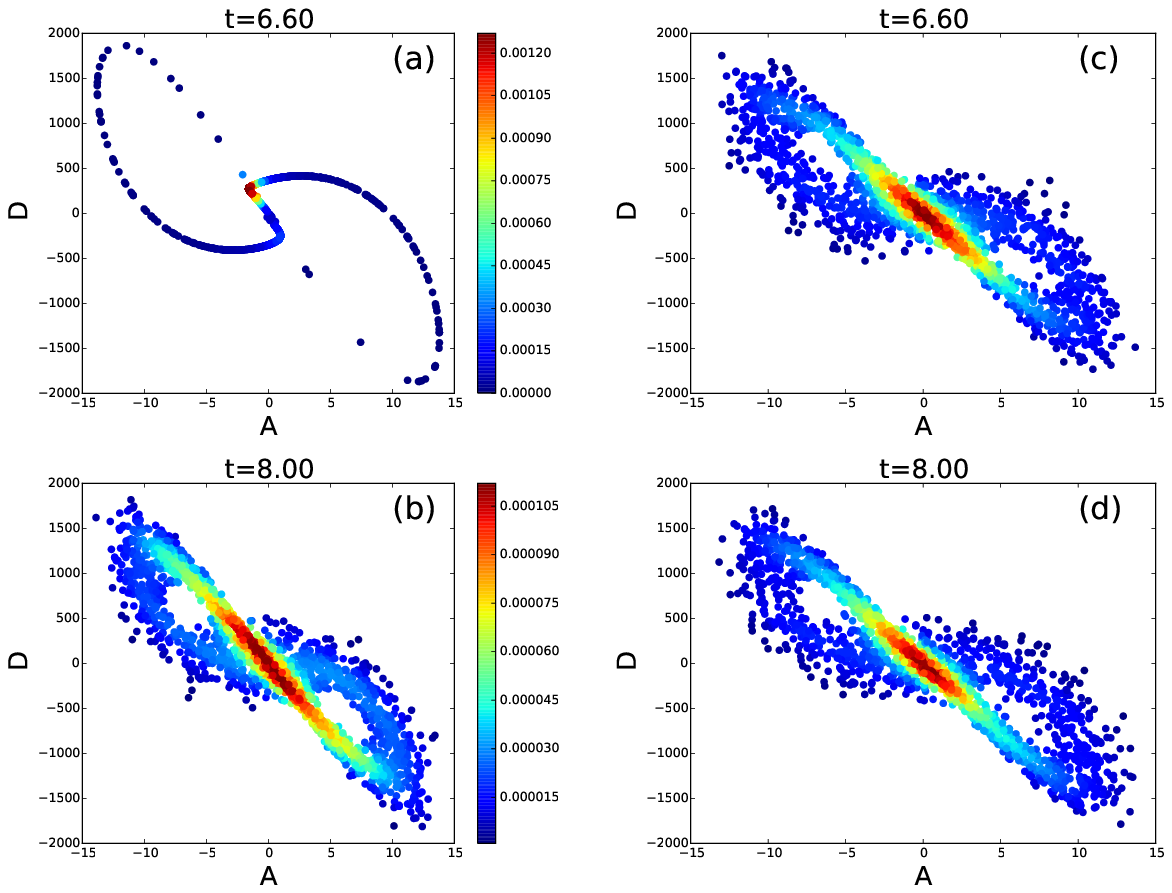}
  \caption{The scatter diagrams of the probability distribution function (PDF) in the $A-D$ plane of the 3-DOF chaotic system (\ref{3eqns})  at different time.  (a) and (b): reliable results given by the CNS using the 90 decimal-digit multiple precision for every data;  (c) and (d): the unreliable results given by double precision.}
\label{fig:3eqns-scatter-round-off}
\end{figure}

The scatter diagrams of the probability distribution function (PDF) in the $A-D$ plane of the 3-DOF chaotic dynamic system (\ref{3eqns})  given by the two numerical schemes (i.e. $M=10$ and $M=80$) are as shown in Figure~\ref{fig:3eqns-GaussianKDE}, where the PDF is obtained by using the Gaussian kernel density estimator.  Note that  the PDFs given by the two numerical approaches are  almost the same at $t=8.00$.  However, at  $t=6.60$ and $t=6.80$,  the PDFs given by $M=10$ are quite different from those given by the CNS using $M=80$.  This supports our previous conclusion that the truncation error has a significant influence on the computation of unsteady statistics, but has no influence on steady ones.

Note that the double precision is widely employed in numerical simulations, which leads to round-off errors at every time-step that increase exponentially for a chaotic dynamic system.   To simulate the round-off error,  we add a random data  at each time-step with zero mean and the standard deviation  $10^{-16}$, while the 80th-order Taylor expansion ($M=80$) is still used and all data are  expressed in 90 decimal-digits so as to guarantee the negligible truncation error in the considered interval of time $t\in[0,10]$.

Figure~\ref{fig:3eqs-M80K300-roundoff} shows the comparison between the reliable CNS statistics (using the 90 decimal-digit multiple precision for every data) and the unreliable results using the double precision.   Note that the round-off error has a great influence on the standard deviation of $A(t)$ from the very beginning.  At about $t \approx 3.5$ when the round-off error enlarges exponentially to reach the level of the ``true'' solution,   the unsteady mean of $A(t)$ becomes unreliable.  Note that the mean of $A(t)$ given by the double precision becomes time-independent more early, at $t \approx 3.5$, which is however wrong in physics.  As the system truly becomes unsteady  when $t>7$ (that is determined by the reliable CNS result),  the  round-off  error  has  no  influence on the computation of statistics.      Furthermore, Figure~\ref{fig:3eqns-scatter-round-off} shows the scatter diagram of the probability distribution function (PDF)  in the $A-D$ plane of the 3-DOF chaotic dynamic system (\ref{3eqns})  given by the two different numerical schemes.    Note that  the PDFs  are almost the same at $t=8.00$.  However, the two PDFs are  obviously different at $t=6.60$.   All of these suggest that the round-off error  has a great influence on the unsteady statistics when the chaotic dynamic system is in a transition stage from an equilibrium state to a new equilibrium one, but has a little impact on steady ones.  

\begin{figure}[h]
  \centering
  \includegraphics[scale=0.45]{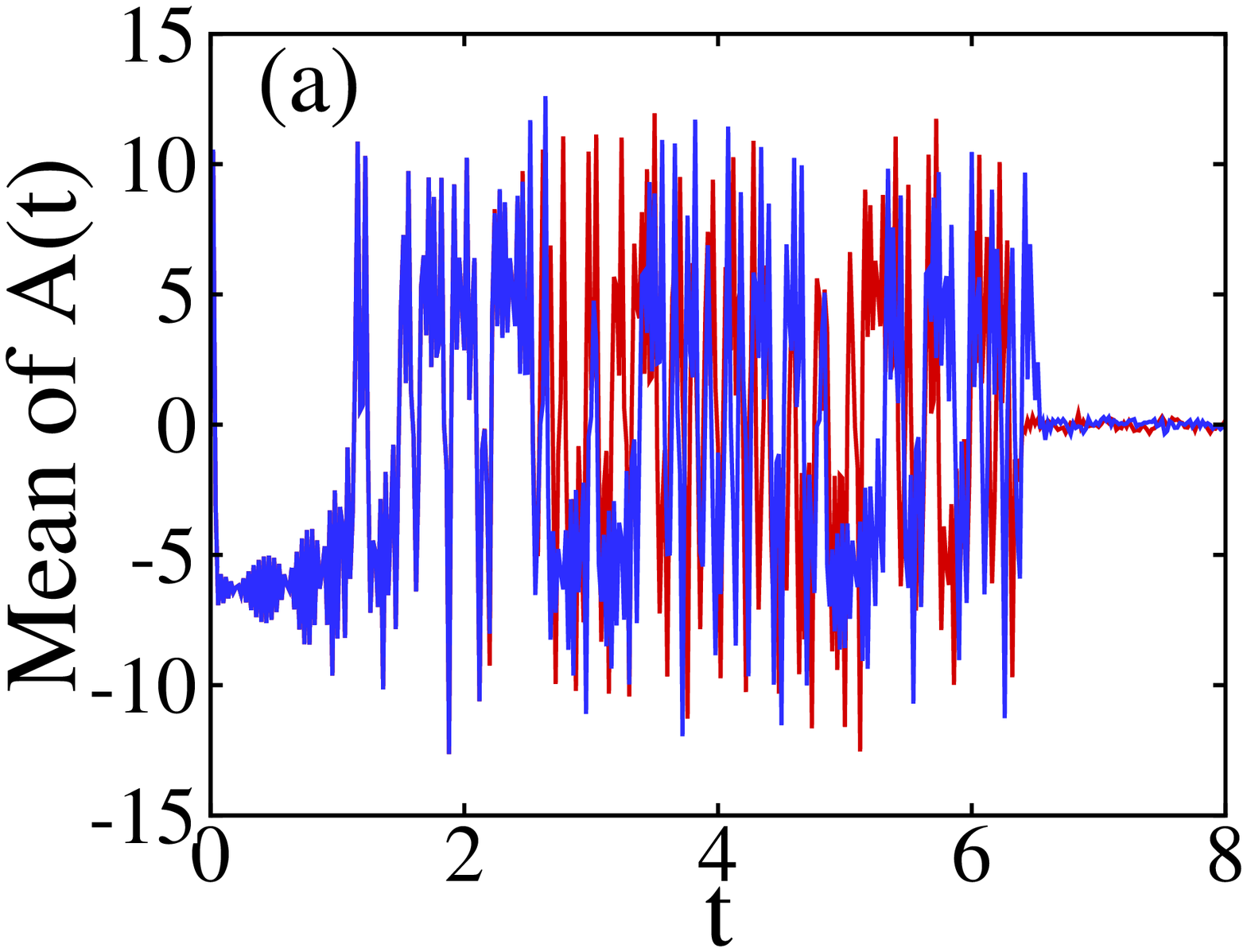}
  
  \includegraphics[scale=0.45]{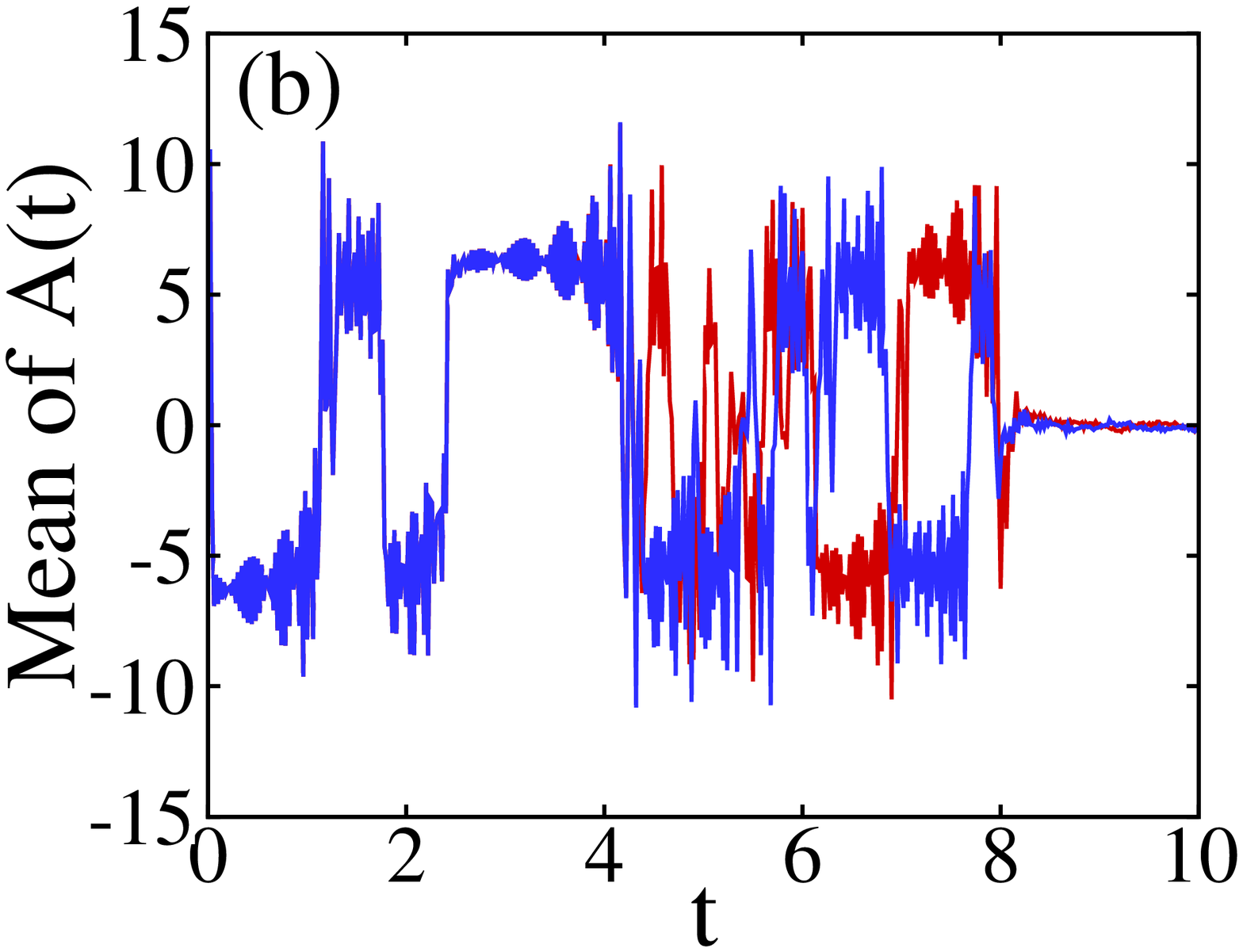}
  \caption{The influence of the truncation error on the mean of $A(t)$ of (a) the 5-DOF  and (b) the 7-DOF chaotic systems obtained using the time-step $\delta t =10^{-3}$  and the 90 decimal-digit multiple precision for every data with the negligible round-off error.   Solid line in red:  the reliable  results given by the CNS using the 80th-order Taylor series expansion for $t$ (i.e. $M=80$);  dashed line in blue:  the unreliable result given by the 10th-order Taylor series expansion (i.e. $M=10$). }
\label{fig:5eqns-7eqns}
\end{figure}

Similarly, we also investigate the influence of numerical noises on computations of statistics of the  5-DOF   and  7-DOF  chaotic dynamic systems,  governed by (\ref{5eqns})  and and (\ref{7eqns}), respectively,  and obtain the same conclusions qualitatively,  as shown in Figure~\ref{fig:5eqns-7eqns}.    It suggests that the above-conclusions are qualitatively the same even for chaotic systems with large numbers of DOF.

\section{Conclusions}

Due to the  the sensitive dependence on the initial condition (SDIC)  (i.e. the famous ``butterfly-effect'') \cite{Lorenz1963},  the numerical noises (such as truncation error and round-off error) enlarge exponentially to reach the same level of the ``true'' solution.  Thus, it is practically very difficult to kick out the exponentially enlarging  numerical noises so as to gain the ``true'' chaotic solutions with physical meanings, and many spurious numerical solutions were reported \cite{Yee1999, Wang2009, Yee1991, Yee1995}.   However,  in practice,  it is widely believed that the numerical noises do not influence the statistic results of chaotic dynamic systems.   In this paper, we compare the statistic results of three chaotic dynamic model equations gained by two different approaches, one is the traditional numerical approach, the other is  the so-called Clean Numerical Simulation (CNS)  with extremely small numerical noises \cite{Liao2009,Liao-Wang2014, Liao2014}.   It is found that the numerical noises have a great influence on the computation of unsteady statistics of chaotic dynamic systems in a transition stage from an equilibrium state to a new equilibrium one, but have no influence on time-independent statistics for systems in an equilibrium state.

It is widely believed that turbulence possesses many characteristics of chaos, such as the sensitive dependence on the initial condition (SDIC), and so on.    This is because turbulent flows can be described by the Navier-Stokes equations, and besides the chaotic model equations considered in this paper are deduced from the N-S equations.  It is indeed true that, when  turbulent flows are on a statistical stationary state,  DNS results often agree well with experimental data, as reported in \cite{Lee2015,KMM1987}.   So,  based on our computations mentioned above,  we have many reasons to believe that the numerical noises should have a great influence on the computation of unsteady statistics of the DNS, but have no influence on the steady ones.   It suggests that we should be extremely careful on the DNS statistic results of unsteady turbulent flows far from an equilibrium state.  

In this paper the CNS is used to gain reliable numerical simulations of chaotic dynamic systems.  Note that the numerical noises of the CNS can be controlled to be much smaller than the ``true'' chaotic solution in a long enough interval of time.  So, the CNS might open a new way to reliably simulate chaotic dynamical systems and even turbulent flows.

\section*{Acknowledgment}
The calculations were performed on TH-1A at National Supercomputer Centre in Tianjin, China.   This work is partly supported by National Natural Science Foundation of China (Approval No. 11272209 and 11432009).

%% The Appendices part is started with the command \appendix;
%% appendix sections are then done as normal sections
%% \appendix

%% \section{}
%% \label{}

%% References
%%
%% Following citation commands can be used in the body text:
%% Usage of \cite is as follows:
%%   \cite{key}         ==>>  [#]
%%   \cite[chap. 2]{key} ==>> [#, chap. 2]
%%

%% References with bibTeX database:

\bibliographystyle{elsarticle-num}
\bibliography{ref}

%% Authors are advised to submit their bibtex database files. They are
%% requested to list a bibtex style file in the manuscript if they do
%% not want to use elsarticle-num.bst.

%% References without bibTeX database:

% \begin{thebibliography}{00}

%% \bibitem must have the following form:
%%   \bibitem{key}...
%%

% \bibitem{}

% \end{thebibliography}
\end{document}